\documentclass[runningheads]{llncs}
\usepackage[T1]{fontenc}
\usepackage[misc,geometry]{ifsym}
\usepackage{graphicx}
\usepackage{cite}
\usepackage{amsmath,amssymb,amsfonts}
\usepackage{algorithmic}
\usepackage{enumitem}
\usepackage{subfigure}
\usepackage{graphicx}
\usepackage{textcomp}
\usepackage{float}
\usepackage{algorithm}
\usepackage{gensymb}
\usepackage{etoolbox}
\usepackage{rotating}
\usepackage{comment}
\usepackage{balance}
\usepackage{booktabs}
\usepackage{hyperref}
\usepackage{CJK} 
\usepackage{tikz}
\usepackage{subcaption}
\usepackage{graphicx}
\usepackage{textcomp}
\usepackage{gensymb}
\usepackage{xcolor}
\usepackage{soul}
\usepackage{hhline}
\usepackage{multirow}
\usepackage{boldline} 
\usepackage{array}
\usepackage{ragged2e}

\newcolumntype{P}[1]{>{\centering\arraybackslash}p{#1}}
\newcolumntype{M}[1]{>{\arraybackslash}m{#1}}
\begin{document}
\title{Evolutionary Optimization of 1D-CNN for Non-contact Respiration Pattern Classification}
\titlerunning{1D-CNN Optimization for Non-contact Respiration Classification}
\author{Md Zobaer Islam\inst{1}\textsuperscript{(\Letter)} \and
Sabit Ekin\inst{2} \and
John F. O'Hara\inst{3} \and
Gary Yen\inst{3}}
\authorrunning{M. Z. Islam et al.}
%
\institute{Department of Radiology, University of North Carolina at Chapel Hill, Chapel Hill, NC, USA\\
\email{zobaer\_islam@med.unc.edu}\\ \and
Departments of Engineering Technology, and Electrical \& Computer Engineering, Texas A\&M University, College Station, TX, USA\\
\email{sabitekin@tamu.edu}\\ \and
Electrical and Computer Engineering, Oklahoma State University, Stillwater, OK, USA \\
\email{\{oharaj, gyen\}@okstate.edu}\\ }
\maketitle              
\begin{abstract}
In this study, we present a deep learning-based approach for time-series respiration data classification. The dataset contains regular breathing patterns as well as various forms of abnormal breathing, obtained through non-contact incoherent light-wave sensing (LWS) technology.  Given the one-dimensional (1D) nature of the data, we employed a 1D convolutional neural network (1D-CNN) for classification purposes. Genetic algorithm was employed to optimize the 1D-CNN architecture to maximize classification accuracy. Addressing the computational complexity associated with training the 1D-CNN across multiple generations, we implemented transfer learning from a pre-trained model. This approach significantly reduced the computational time required for training, thereby enhancing the efficiency of the optimization process.  This study contributes valuable insights into the potential applications of deep learning methodologies for enhancing respiratory anomaly detection through precise and efficient respiration classification.

\end{abstract}
\section{Introduction}
Human respiratory rate and its pattern convey valuable information about the physical and mental states of the subject. Anomalous breathing can be indicative of serious health issues such as asthma, bronchitis, emphysema, lung cancer, COVID-19, etc. It can also serve as a precursor to unstable mental conditions, including stress, panic, anxiety, etc. Contact-based monitoring falls short in capturing the true essence of one's breathing, as individuals may alter their breathing patterns when aware of being monitored. Furthermore, subjects may be too young or too ill to apply contact-based approaches for an extended duration. Current automated technologies for non-contact breathing monitoring typically rely on sensing electromagnetic signals, such as radar, WiFi, or extracting breathing information from videos captured by red-green-blue (RGB) or thermal infrared cameras~\cite{Purnomo2021, Gu2019, Brieva2020, Jagadev2020, Jakkaew2020}. We have developed a non-contact respiration monitoring system based on sensing incoherent infrared (IR) light reflected from the subject's chest. This technology surpasses existing methods due to the safe, ubiquitous, and discreet nature of infrared light, avoiding privacy issues associated with camera-based approaches. The main functional components of the system include the LWS hardware, signal processing, and feature extraction algorithms, along with classification algorithms. Infrared light-emitting diode (LED) sources illuminate the subject's chest, and the reflected light's intensity, varying with chest movement, is captured and converted into electrical voltage by a commercially available photodetector. Collected data are processed to extract features using signal processing algorithms and then classified using machine learning models.

Classifying 1D respiration data for respiratory anomaly detection is a crucial task in healthcare. Traditional machine learning models, including decision tree, random forest, K-nearest neighbor, and support vector machine, have been commonly employed for this purpose. However, these models heavily rely on handcrafted features, necessitating domain expertise and environment-specific additional measurements. This dependence can be considered an overhead in live implementation~\cite{islam2023}. In contrast, deep learning models, such as 1D-CNNs, have shown promise in automatically extracting subtle features, including those imperceptible to humans. Leveraging convolutional layers for feature extraction and fully connected layers for classification, 1D-CNNs offer a more suitable approach for real-life implementation. Consequently, this study explores the application of 1D-CNNs for the classification of respiration data.

\begin{figure}[!htbp]
 \centering \includegraphics[width=0.6\textwidth]{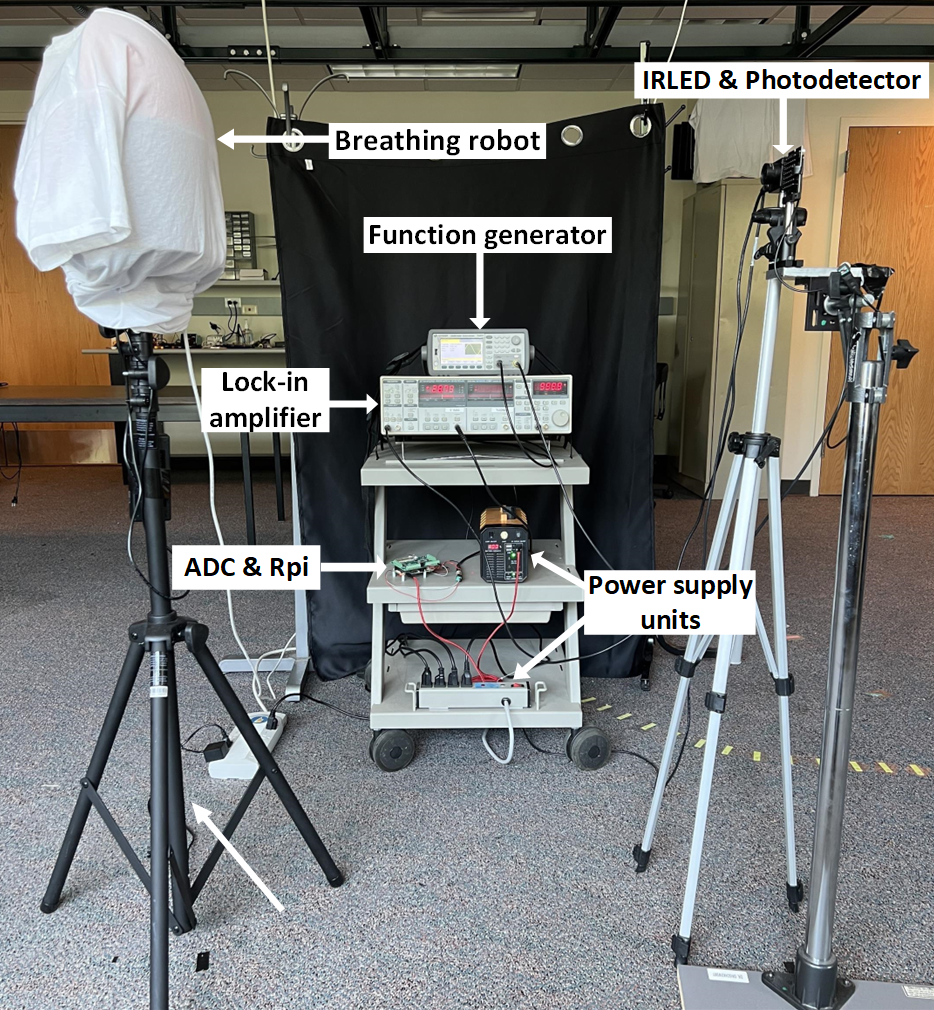}
 \caption{The hardware setup of the overall system used for data collection.}
 \vspace{-4.3mm}
 \label{fig:setup2}
 \end{figure}
 
Neural architecture search, a prominent research area, aims to identify the most optimal neural network architecture for specific tasks. While random search is not always guaranteed to yield an optimal model, it is not uncommon in the literature~\cite{ragab2020random,hsieh2020detection}. Genetic algorithm (GA)~\cite{katoch2021review}, renowned for single and multi-objective optimization tasks, has shown promise in neural architecture search~\cite{lu2019nsga,xiao2020efficient}. Various GA variants have been utilized for CNN optimization in diverse applications, such as fault diagnosis of hydraulic piston pumps, Alzheimer’s disease diagnosis, and stock market prediction~\cite{ugli2023automatic,lee2021genetic,chung2020genetic}. In the current study, 1D-CNN is employed to classify respiration data into 8 different classes, including a class dedicated to faulty data detection. The architecture of the 1D-CNN is optimized to maximize test accuracy through a genetic algorithm developed from scratch in Python. To minimize computational complexity, transfer learning is employed from a pre-trained 1D-CNN model, enabling efficient execution of the genetic algorithm across multiple generations.

The remainder of this manuscript is organized as follows. Section~\ref{sec:System_Design} describes the system model including  hardware setup and software algorithms to collect breathing data of different classes and perform classification. Section~\ref{sec:1dcnnopt} presents the details of 1D-CNN optimization including the application of transfer learning and genetic algorithm. Next, Section~\ref{sec:Results} presents the results and discussion from GA and final breathing classification using the obtained optimized model. Finally, Section~\ref{sec:Conclusion} draws conclusions from the effort and forecasts future research directions.

\section{System Design and Implementation}
\label{sec:System_Design}

\subsection{Experimental setup}
Human subjects typically cannot replicate breathing with precise frequency, amplitude, and patterns consistently. Therefore, a mechanical robot capable of simulating human breathing in various programmable patterns was developed and utilized for this study. The LWS sensing hardware includes infrared light sources as transmitters, a photodetector as the receiver, a digital signal processing (DSP) unit to convert the received signal from analog to digital, and an electronic module for processing and storing the digital data. Additionally, a function generator and lock-in amplifier were used to for coherent detection of the received light. The light source utilized was an invisible 940\,nm IR lamp board (48 black LED illuminator array), offering a range of 30\,ft and a wide beamwidth of 120\degree~wide beamwidth. The light source was modulated by a 1\,kHz sinusoidal voltage wave from a function generator. The receiver employed was a commercial photodetector, Thorlabs PDA100A with a converging lens. For lock-in detection, SR830DSP frequency lock-in amplifier was utilized. A Raspberry Pi, equipped with an analog-to-digital converter (ADC) circuit, managed the data collection, digitization, and storage for subsequent offline processing. The complete experimental setup is illustrated in Fig.~\ref{fig:setup2}.

\begin{table}[!htbp]
\renewcommand{\arraystretch}{1.15}
\centering
\vspace{-6mm}
\caption{Characteristics and number of collected data of each breathing class}
\begin{center}
\begin{tabular}{|P{1.1cm}|M{2cm}|P{2cm}|P{2cm}|P{2.5cm}|} 
 \hline
\bf{Class} & \bf{Class name} &  {\centering \bf{Breathing rate (BPM)}} &  {\centering \bf{Breathing depth (\%)}} & {\centering \bf{Number of data instances}}\\
 \hline
  \bf{0} & Eupnea & 12-20 & 30-58 & 300\\
  \bf{1} & Apnea & 0 & 0 & 300\\
  \bf{2} & Tachypnea & 21-50 & 30-58 & 300\\
  \bf{3} & Bradypnea & 1-11 & 30-58 & 300\\
   \bf{4} & Hyperpnea & 12-20 & 59-100 & 300\\
   \bf{5} &Hypopnea & 12-20 & 1-29 & 300\\
   \bf{6} &Kussmaul's & 21-50 & 59-100 & 300\\
   \bf{7} &Faulty data &Any &Any & 300\\
     \hline
      \multicolumn{4}{|r|}{\RaggedRight \bf{Total}} & 2400\\
 \hline
\end{tabular}
\end{center}
\vspace{-8.5mm}
\label{table:classes}
\end{table}

\subsection{Data collection}
To gather respiration data, the infrared light emitted by the LED source underwent modulation using a sinusoidal voltage wave of 1 kHz frequency from the function generator. The reflected light was captured by a photodetector paired with a converging lens. The photodetector's gain was set to 40 dB. The output from the photodetector was linked to a lock-in amplifier with a time constant of 100 ms. The resulting filtered signal from the lock-in amplifier was then digitized by the ADC. Finally, a Python script executed on the Raspberry Pi collected and stored the voltage amplitude data, along with their corresponding timestamps.

\begin{figure}[t]
    \centering
    \includegraphics[width=.94\textwidth,height=.6\textwidth]{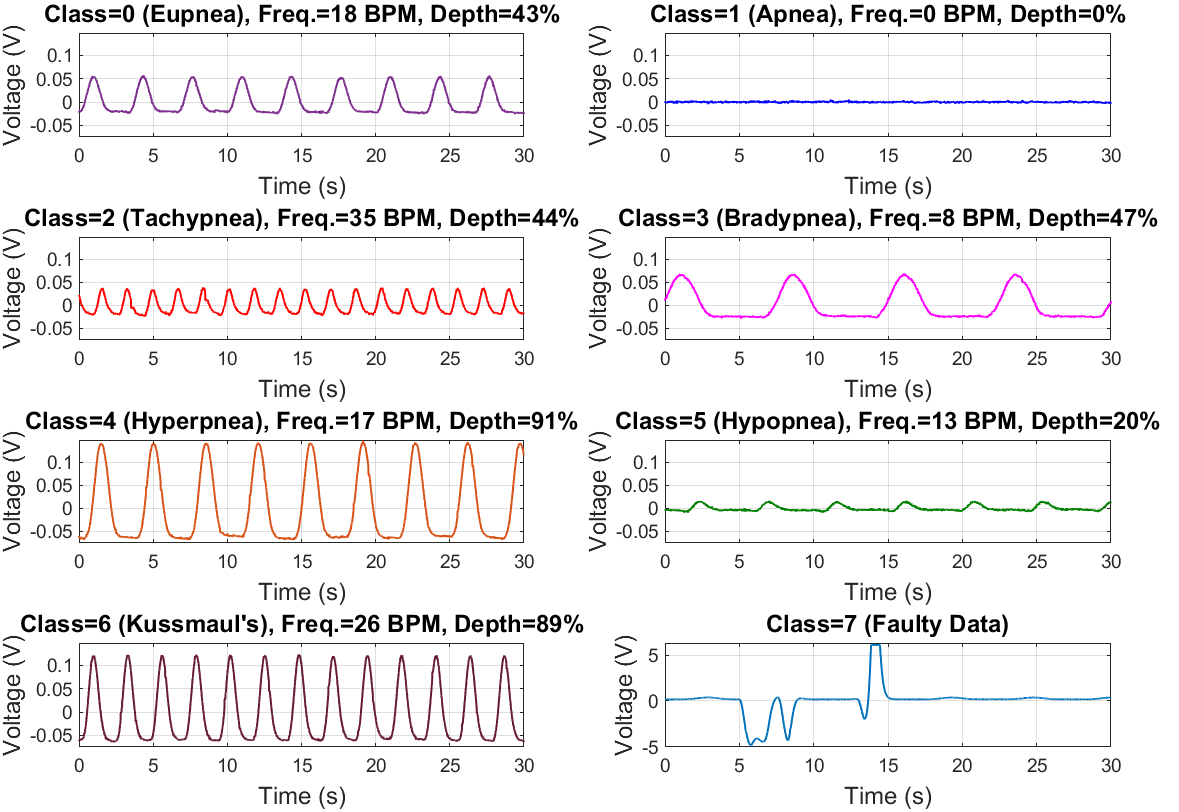}
    \caption{Time domain representation of sample data from each class (the distance between the source-photodetector and the robot was 1\,m).}
    \vspace{-3.6mm}
    \label{fig:data_visual}
\end{figure}

From the literature, 7 different classes of human breathing were identified and their characteristics, in terms of breathing rate (frequency) and depth (amplitude)\footnote{Breathing depth is measured from the human rib cage movement and it is expressed as a percentage of the maximum movement of the rib cage~\cite{Parreira2010}.}, were summarized in Table~\ref{table:classes}~\cite{BarbosaPereira2017, Purnomo2021, Parreira2010,Ali2021,Rehman2021,Jagadev2020,Moraes2019,Fekr2016,Ucar2017}. One of them (Eupnea) was normal breathing, while the rest were examples of different categories of abnormal breathing.  The robot was programmed to breathe with frequencies and amplitudes in the appropriate ranges according to Table I to generate data experimentally for each class. Data for one additional class called `faulty data' were generated too by manually interrupting the system or the surroundings during data collection. This class was introduced to detect erroneous data generated due to possible internal or external malfunction so that they can be discarded and recollected. $\sin^6$ waveform seemed to have the closest match with human breathing in terms of inspiration and expiration time, hence this pattern was used in the robot for all breathing classes. A total of 2400 data instances (300 per class)  were collected during the daytime, keeping the windows of the room unshaded and internal lighting common for an office environment. For each class, 100 data instances were collected at 0.5\,m, 1\,m and 1.5\,m distances each. Each data was 30\,s long and collected at 100\,Hz sampling frequency. All the collected data were saved along with their class labels (0 to 7) for further offline processing. Fig.~\ref{fig:data_visual} shows the time domain waveforms of a few samples of raw data, one from each class, collected at 1\,m distance.

The raw data went through offline signal processing which includes moving average filtering (50 points) to suppress noise and a 5-th order polynomial detrending to remove upward or downward trends that were in some data instances. Thus, the data were prepared for feeding them to 1-dimensional convolutional neural network (CNN) for feature extraction and classification.

\section{1D-CNN Optimization}
\label{sec:1dcnnopt}
Given that the respiratory data is one-dimensional, the application of a 1D-CNN with one-dimensional filters for convolution becomes a viable choice. Leveraging the Keras API in Python, the data is processed through a 1D CNN model. The 1D-CNN model may have various potential architectures, each resulting in different accuracies for respiration classification. To optimize accuracy, it becomes imperative to identify the most suitable architecture. The utilization of a genetic algorithm (GA) is proposed to fine-tune the parameters of the 1D-CNN model, aiming to achieve an optimal architecture. However, the evaluation of fitness values for numerous solutions or 1D-CNN architectures is computationally expensive. To address this challenge, transfer learning has been incorporated into this study. By leveraging transfer learning, the computational burden is significantly reduced, enabling the execution of the genetic algorithm over a sufficient number of generations.

\subsection{Transfer Learning Application}
\label{subsec:tf}
Transfer learning is a powerful paradigm in machine learning that leverages knowledge gained from solving one task to improve the performance on a different, but related, task. In essence, it allows a model trained on one dataset to be fine-tuned or adapted for a different but similar task. This approach is particularly beneficial when the target task has limited labeled data, as it enables the model to generalize better. It is also useful to make the training faster through utilizing pre-trained weights. In the context of deep learning, transfer learning is often applied using pre-trained neural network architectures, such as those trained on large-scale image datasets like ImageNet. The model's initial layers learn generic features like edges and textures, while the deeper layers capture more task-specific information. This modular structure facilitates the reuse of the pre-trained layers for a new task. For instance, in Python, utilizing transfer learning with TensorFlow library involves loading pre-trained models like VGG16 or ResNet and adapting them for specific tasks with minimal effort. This accelerates training and enhances performance, making transfer learning a valuable tool in the machine learning toolbox.

\begin{figure}[!htbp]
\vspace{-4.6mm}
\includegraphics[width=0.89\textwidth]{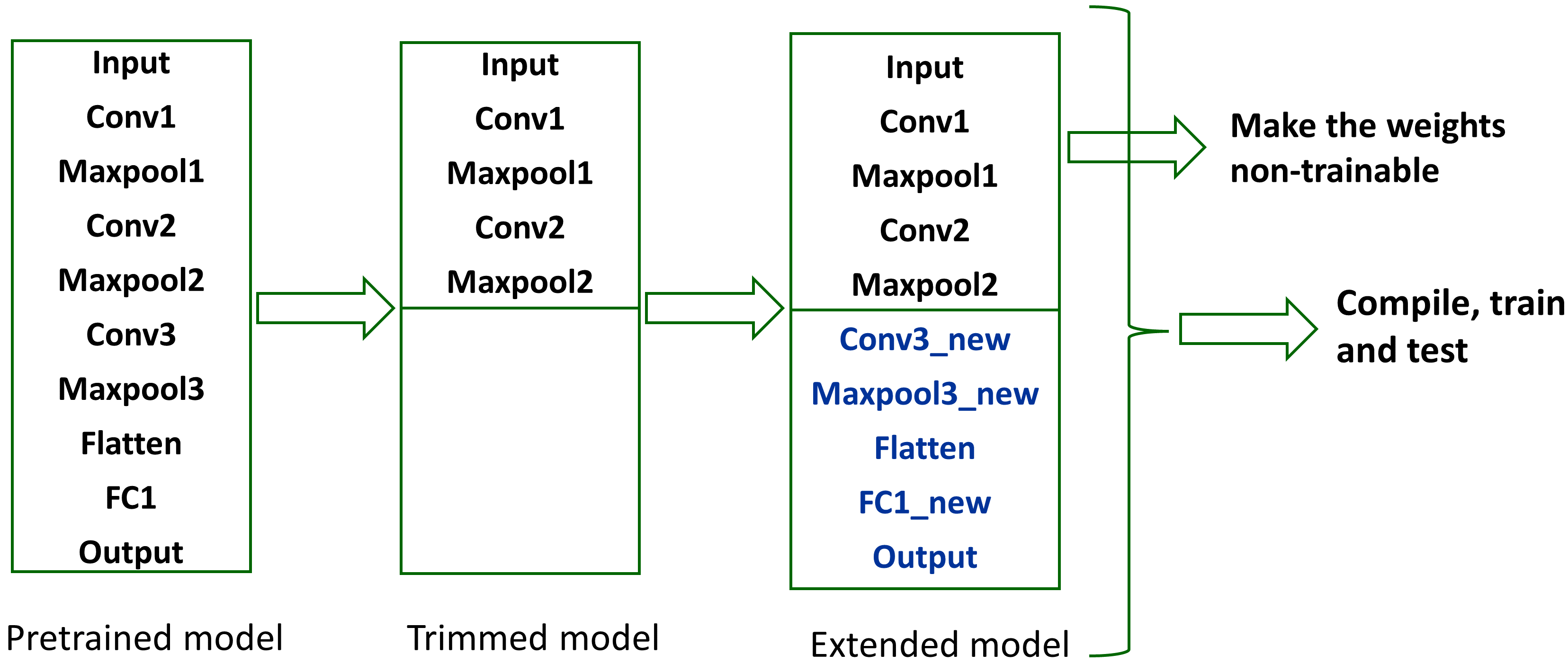} 
\centering
\caption{Pre-trained, trimmed and extended models.}
\vspace{-4.6mm}
\label{fig:tf_model}
\end{figure}

Although many famous 2D-CNN models pre-trained on ImageNet dataset are available as API for direct use, their 1D counterparts are rare in the literature and GitHub repositories. To address this problem, a single 1D-CNN model was obtained through random search that yields higher classification accuracy than most of the other tested models. The complete architecture of that model is shown in the leftmost bounded box in Fig.~\ref{fig:tf_model} and described as follows:
\begin{enumerate}
    \item \textbf{Input layer}: It was a breathing data vector (1, 3000) pulled from the rows of the overall data matrix.
\item \textbf{Convolutional layer 1}: It was the first convolutional layer. 256 1D filters of length 64 were used to perform convolution in this layer. Spatial dimensions of data were preserved using padding =``same''. `Relu' was used as the activation function.
\item \textbf{Maxpooling layer 1}: This layer took 2 data points repeatedly from the previous layer and replaced it by the maximum value. So, it had length 2 and stride 2.
\item \textbf{Convolutional layer 2}: It was the second convolutional layer. 128 1D filters of length 32 were used to perform convolution in this layer. Similar padding and activation function was used as the first convolutional layer.
\item \textbf{Maxpooling layer 2}: This layer is exactly the same as the first maxpooling layer
\item \textbf{Convolutional layer 3}: It was the third convolutional layer. 64 1D filters of length 16 were used to perform convolution in this layer. Similar padding and activation function was used as the first and second convolutional layers.
\item \textbf{Maxpooling layer 3}: This layer is exactly the same as the first and second maxpooling layers.
\item \textbf{Flattened layer}: The resulting data were flattened for feeding it to the fully connected (FC) layer.
\item \textbf{Fully connected layer 1}: The flattened data was passed through a fully connected neural layer. The hidden layer had 64 neurons. `Relu' was used as the activation function.
\item \textbf{Output layer}: It was the final output layer with `softmax' activation function which had 8 neurons, one for each class.

\end{enumerate}

 \begin{figure*}[!htbp]
 \vspace{-3.6mm}
\includegraphics[width=\textwidth]{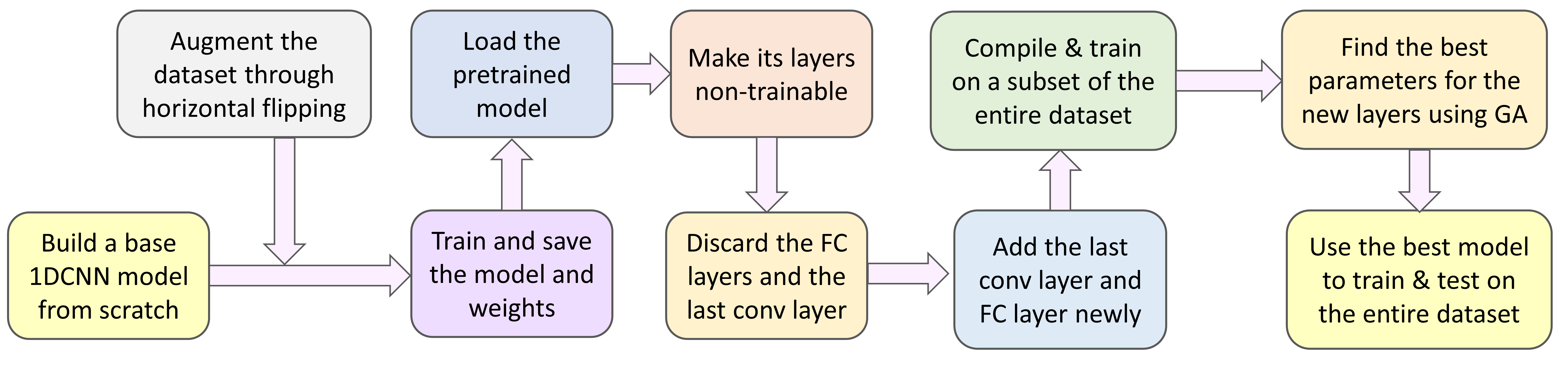} 
\centering
\caption{Transfer learning framework for 1-dimensional respiration data classification.}
\vspace{-4.6mm}
\label{fig:tf}
\end{figure*}

 \begin{figure}[!htbp]
  \vspace{-3.6mm}
\includegraphics[width=0.98\textwidth]{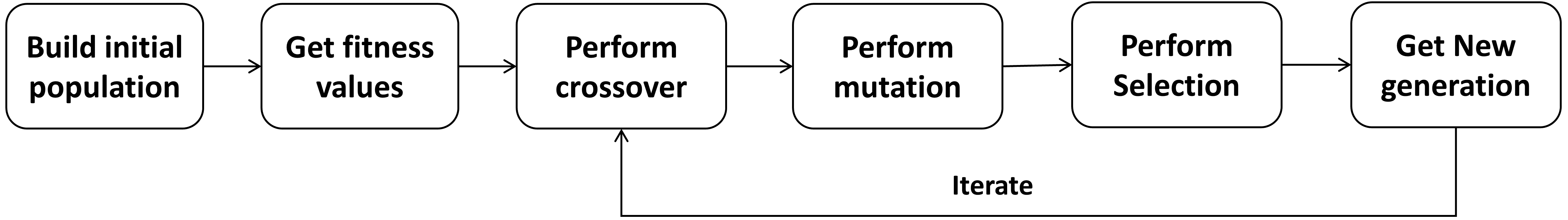} 
\centering
\caption{Genetic algorithm workflow.}
 \vspace{-4.6mm}
\label{fig:ga}
\end{figure}
Next, this model was trained for 30 epochs and the trained model and weights were stored as pre-trained model. It is common in transfer learning to apply data augmentation to increase the number of training data and improve generalization. For image dataset, various affine transformations like rotation, shearing etc. are applied to produce new data instances~\cite{taylor2018improving,montserrat2017training}. In our case, we not only utilized the whole respiration dataset of 2400 data instances, but also doubled the number of data through horizontal flipping. Since the respiration simulated by the robot was symmetric, following a $\text{sin}^6$ pattern, horizontal flipping generated valid respiration data. Then the combined 4800 instances of data (with 80\%-20\% train-test split) were fed to the 1D-CNN model for training. After that, the pre-trained model and weights were saved in .h5 format. The overall approach followed for transfer learning part is presented in Fig.~\ref{fig:tf}.

\begin{figure}[t]
\includegraphics[width=0.96\textwidth,height=0.142\textheight]{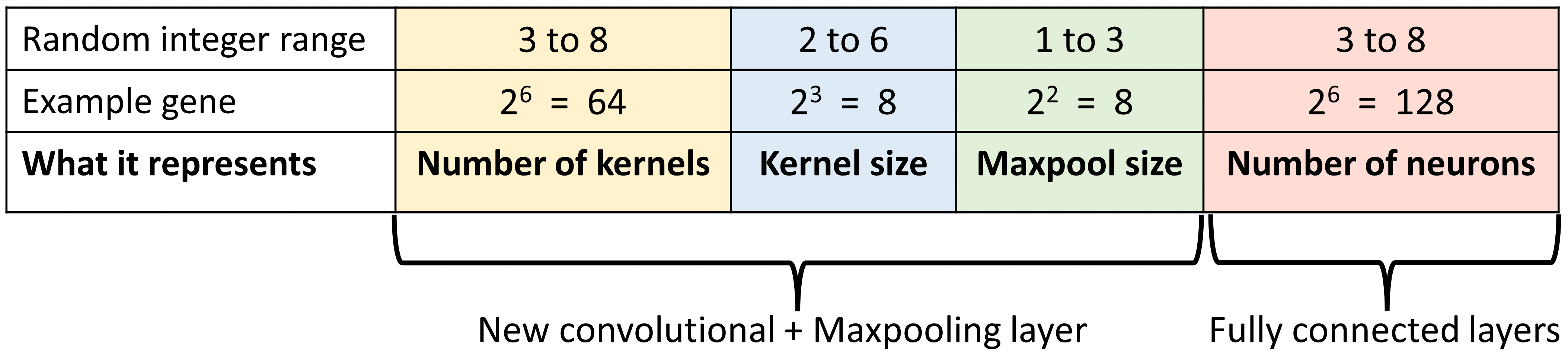} 
\centering
\caption{Chromosome for GA.}
 \vspace{-4mm}
\label{fig:chrom}
\end{figure}

\subsection{Optimization Using Genetic Algorithm}
\label{subsec:ga}
Genetic algorithm was applied using Python to find the optimum 1D-CNN model from a chosen solution space through crossovers and mutations over generations of solutions. Since transfer learning was applied to preload part of the CNN model in trained status, GA was used mainly to tune the rest of the model. The high-level workflow of the genetic algorithm applied was shown in Fig.~\ref{fig:ga}, but it involved some advanced features too which will be discussed here.

At first, the saved pre-trained model was loaded. Then, as depicted in Fig.~\ref{fig:tf_model}, the output layer, FC layer, flattened layer and the last convolutional and maxpooling layers were removed. After that, similar untrained layers were added newly to construct the extended model (Fig.~\ref{fig:tf_model}). The weights of the pre-trained part were made non-trainable, because these should not change during the new training runs. Next, GA was applied to choose parameters for the untrained part to maximize the classification accuracy. A subset of the entire data set (1000 data instances with 80\%-20\% train-test split) was used for applying GA to make the training processes faster.

\subsubsection{\textbf{Build Chromosome and Initial Population}:}
The first step was to decide the chromosome structure or genotypes of each gene and define their corresponding phenotypes or how they relate to the 1D-CNN architecture. Our chromosome contains four genes as shown in Fig.~\ref{fig:chrom}. We used all parameters as powers of 2 because these numbers create more computationally efficient model. First gene denotes the number of kernels for the newly added convolutional layer and is a random integer number between 3 to 8. If the integer generated is 6, then the number of kernels for that layer will be $2^6=64$. Second gene denotes the length of kernels and is an integer between 2 to 6, or equivalently kernel length varied between $2^3$ and $2^6$. The third gene represents the maxpooling size and the fourth gene indicates the number of neurons present in the fully connected layer.

\subsubsection{\textbf{Get Fitness Values}:}
In every iterations of GA, new parameters for the untrained part were chosen for each solutions, and the combined model was compiled, trained on training dataset and finally tested on test dataset to get the fitness value. Only one epoch was used for each solution, with a batch size 50, to save computational time. The obtained test accuracy is defined to be the reproductive fitness value.

\subsubsection{\textbf{Perform Crossover}:}
3 single-point crossovers were performed on the chromosomes each having crossover probability 0.8 in each generation. We tried two approaches to select parents for crossover operations. Firstly, the highest fitness 6 individuals were chosen who went through crossover to produce offsprings. Secondly, a Roulette Wheel selection technique was adopted to select parents so that all individuals got at least some chance to be selected.

\begin{figure}[t]
\includegraphics[width=0.82\textwidth]{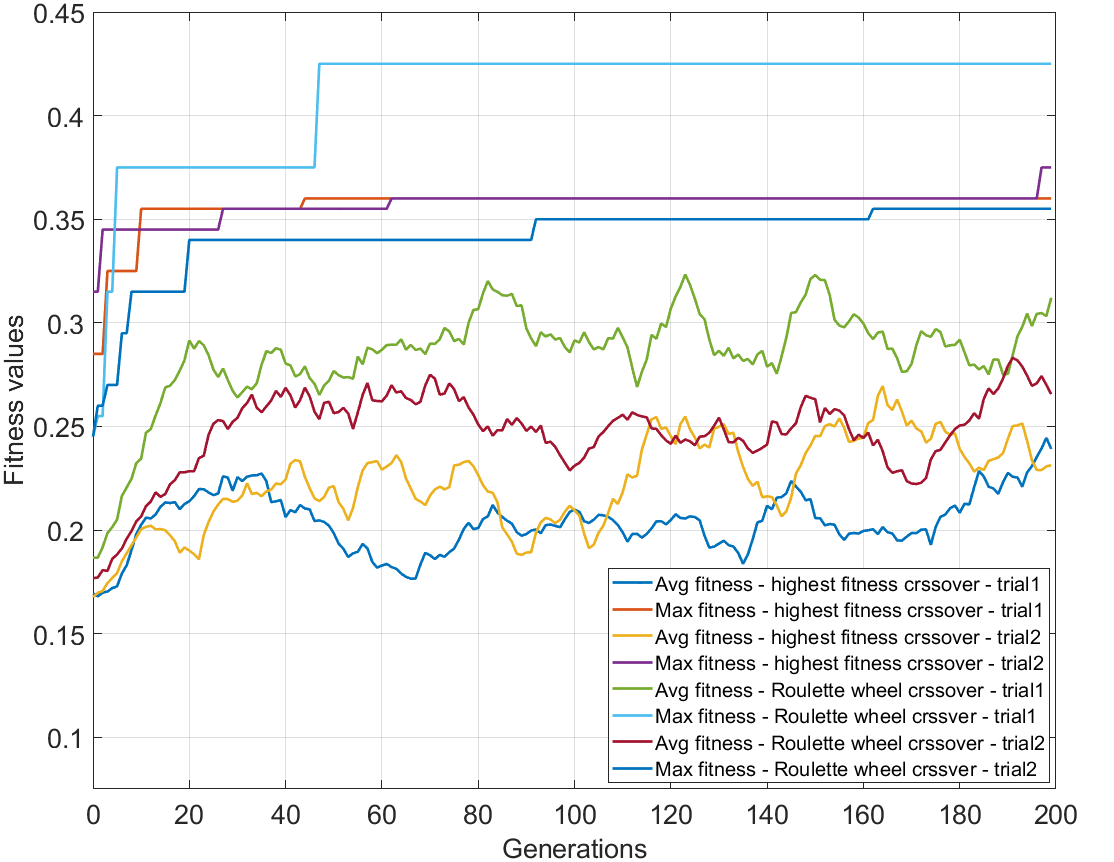} 
\centering
\caption{Maximum and average classification accuracy on the chosen subset of data after one epoch (defined as reproductive fitness) vs. number of generations of solutions}
 \vspace{-3mm}
\label{fig:generations_all_cases}
\end{figure}

\subsubsection{\textbf{Perform Mutation}:}
Frequent mutation might prohibit convergence by losing the best solutions, hence only 1 individual 
went through mutation in each generation with a lower probability (0.4). In each mutation operation, only 1 gene was altered in the chromosome.

\subsubsection{\textbf{Perform Selection}:}
Roulette wheel selection was used to select 8 individuals from the combined list of individuals of current and previous generations. While doing this, 2 best fitness individuals were preserved through elitism. If there were any duplicate chromosomes in the selected individuals, then they were replaced by randomly generated new chromosomes.

\begin{figure}[t]
\includegraphics[width=0.78\textwidth]{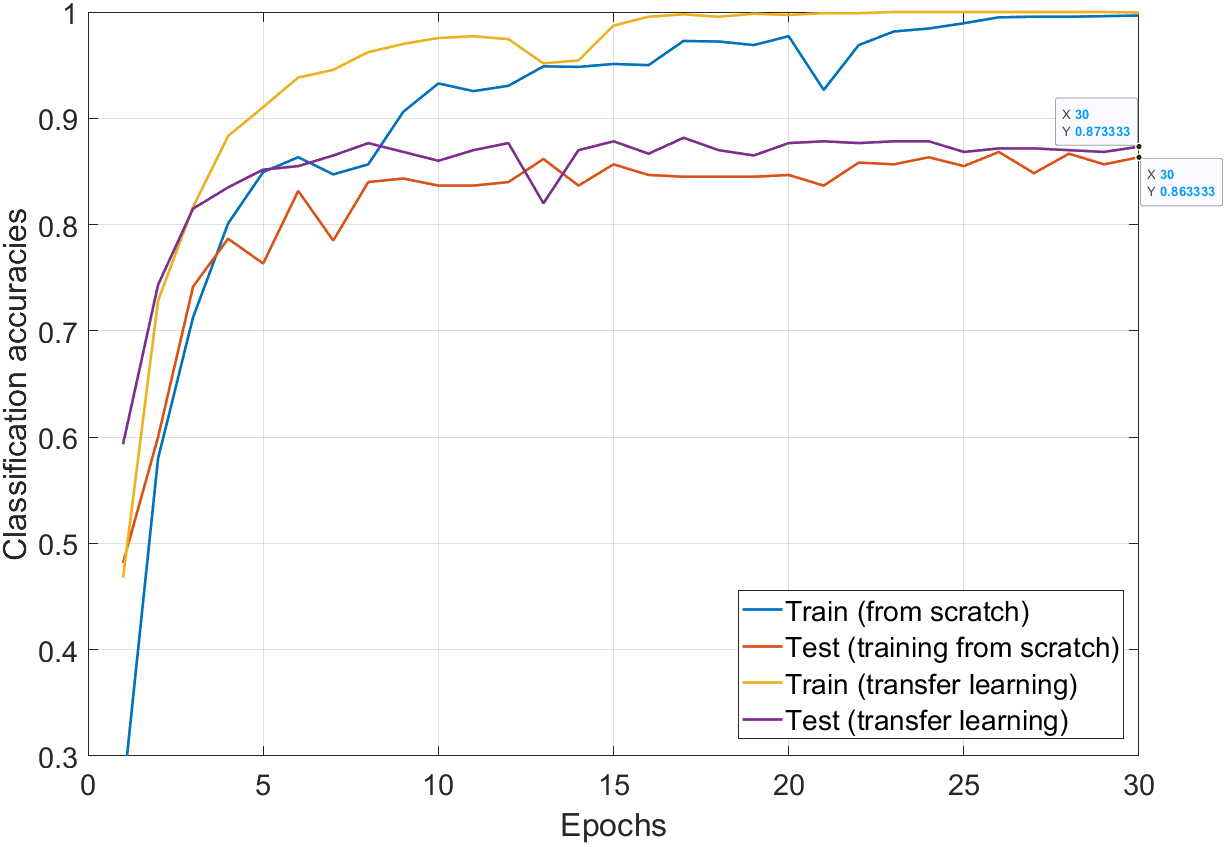} 
\centering
\caption{Training and testing accuracy plotted against the number of epochs. The training was performed by feeding the entire breathing dataset to the optimized 1D-CNN model found through GA.}
 \vspace{-3mm}
\label{fig:final_training}
\end{figure}

\section{Results and Discussion}
\label{sec:Results}
The results from a few trial runs of GA up to 200 generations are displayed in Fig.~\ref{fig:generations_all_cases}. It shows how the maximum fitness and the average fitness evolves over generations. 2 trials are shown with highest fitness individuals selected as crossover parents, while 2 more are shown with Roulette wheel selection for selecting parents for crossover. Average fitness values fluctuate frequently, hence 10 points moving average filtering was performed to see the trend. Although the average fitness values are slowly increasing, there is clear increasing trend in the maximum fitness values.  Trial 1 with the Roulette wheel parents selection for crossover yielded the highest fitness value 0.43 or 43\% classification accuracy after 1 epoch. The corresponding chromosome was [7, 5, 1, 8] and this was chosen as the best solution to test on the entire dataset over many epochs. This solution was combined with the pre-trained part of the model and then the entire model was trained on the available 2400 data instances (without augmentation this time) for 30 epochs. Two distinct training approaches were employed: one involving training the model from scratch with no pre-trained weights, and the other incorporating transfer learning. The latter utilized pre-trained weights for the initial layers, while training only the remaining layers. The training and test accuracies as a function of number of epochs for both scenarios are plotted in Fig.~\ref{fig:final_training}. Notably, the achieved test accuracies were comparable in both training methods, with a slightly higher accuracy (87.33\%, compared to 86.33\%) observed when incorporating transfer learning in the final training.

\section{Conclusion and Future Directions}
\label{sec:Conclusion}

In conclusion, deep learning model like 1D-CNN was found to be effective in respiration data classification which paves the way to respiratory anomaly detection and improves health monitoring at both home and clinics. To further alleviate computational complexity, there is an opportunity to execute the genetic algorithm code exclusively on GPU, leveraging its parallel processing capabilities. More sophisticated data augmentation can be performed beyond horizontal flipping to enhance the dataset for base model training further which will improve the pre-trained weights. The problem can be made multi-objective by including a second objective of minimizing the number of trainable parameters in the 1D-CNN architecture and non-dominated sorting genetic algorithm (NSGAII) can be applied to find the optimum model.

\subsubsection{Acknowledgements} This work was supported in part by the U.S. National Science Foundation under Grants 2008556, 2323301 and 2336852.

%
%
%
\bibliographystyle{splncs04}
%
\bibliography{IEEEexample}

\end{document}